\documentclass[twocolumn]{aastex63}
\usepackage{amsmath,amssymb,graphicx,natbib}
\hypersetup{linkcolor=red,citecolor=blue,filecolor=cyan,urlcolor=magenta}
\shorttitle{KISSR\,102}
\shortauthors{Kharb et al.}
\begin{document}
\title{The Intriguing Parsec-Scale Radio Structure in the ``Offset AGN'' KISSR\,102}

\correspondingauthor{P. Kharb}
\email{kharb@ncra.tifr.res.in}
\author[0000-0003-3203-1613]{P. Kharb}
\affil{National Centre for Radio Astrophysics (NCRA) - Tata Institute of Fundamental Research (TIFR), 
S. P. Pune University Campus, Post Bag 3, Ganeshkhind, Pune 411007, India}
\author[0000-0003-4184-6152]{D. Lena}
\affil{SRON, the Netherlands Institute for Space Research, The Netherlands}
\affil{Department of Astrophysics/IMAPP, Radboud University Nijmegen, PO Box 9010, NL-6500 GL Nijmegen, the Netherlands}
\author[0000-0002-5195-335X]{Z. Paragi}
\affil{Joint Institute for VLBI ERIC, Postbus 2, 7990 AA Dwingeloo, The Netherlands}
\author[0000-0002-5331-6098]{S. Subramanian}
\affil{Indian Institute of Astrophysics, II Block, Koramangala, Bangalore 560034, India}
\author[0000-0003-3295-6595]{S. Vaddi}
\affil{National Centre for Radio Astrophysics (NCRA) - Tata Institute of Fundamental Research (TIFR), S. P. Pune University Campus, Post Bag 3, Ganeshkhind, Pune 411007, India}
\author[0000-0001-8996-6474]{M. Das}
\affil{Indian Institute of Astrophysics, II Block, Koramangala, Bangalore 560034, India}
\author[0000-0001-5574-5104]{Rubinur K.}
\affil{National Centre for Radio Astrophysics (NCRA) - Tata Institute of Fundamental Research (TIFR), S. P. Pune University Campus, Post Bag 3, Ganeshkhind, Pune 411007, India}

\begin{abstract}
We report the detection of an intriguing parsec-scale radio source in the ``offset AGN'' {candidate}, KISSR\,102. The elliptical host galaxy includes two optical nuclei at a projected separation of 1.54~kpc, N1 and N2, to the south-east and north-west, respectively. Phase-referenced VLBA observations at 1.5 and 4.9 GHz of this LINER galaxy, have detected double radio components (A and B) at a projected separation of 4.8~parsec at 1.5~GHz, and another partially-resolved double radio structure at 4.9~GHz coincident with the brighter radio component A. These radio detections are confined to the optical nucleus N1. The brightness temperatures of all the detected radio components are high, $\gtrsim10^8$~K, consistent with them being components of a radio AGN. The $1.5-4.9$~GHz spectral index is inverted {($\alpha\sim+0.64\pm0.08$)} for component A and steep for component B ($\alpha \lesssim-1.6$). The dramatic change in the spectral indices of A and B is inconsistent with it being a typical ``core-jet'' structure from a single AGN, or the mini-lobes of a compact symmetric object. To be consistent with a ``core-jet'' structure, the jet in KISSR\,102 would need to be undergoing strong jet-medium interaction with dense surrounding media resulting in a drastic spectral steepening of the jet. Alternatively, the results could be consistent with the presence of a parsec-scale binary radio AGN, which is the end result of a three-body interaction involving three supermassive black holes in the center of KISSR\,102.
\end{abstract}

\section{Introduction} \label{sec:intro}
All massive galaxies are {believed} to host supermassive black holes in their centres \citep{Kormendy95,FerrareseFordRev05}. Hierarchical galaxy formation models therefore imply the presence of two or more supermassive black holes at the centres of galaxy merger remnants \citep{Volonteri03,Komossa12}. When these black holes eventually coalesce, they emit gravitational radiation and leave behind even more massive remnant black holes \citep{Centrella10,Burke-Spolaor18}. If a triplet of black holes is initially present, three-body interactions would facilitate the coalescence of the inner binary \citep{Hoffman07,Bonetti16}. Alternatively, in a system with two black holes initially, the merged black hole could ``recoil'' due to the asymmetric release of gravitational radiation \citep[e.g.][]{Bekenstein73,Campanelli07}. This could result in an active galactic nucleus (AGN) being ``offset'' from its host galaxy centre \citep[e.g.][]{gualam,BlechaL08,lenaRMA14}. Here we report an intriguing parsec-scale radio structure in an ``offset AGN'' candidate, viz., KISSR\,102. 

This source belongs to the KPNO Internal Spectroscopic Red Survey (KISSR) of emission line galaxies \citep{Salzer00}. It was one among several low ionisation nuclear emission line region (LINER) and Seyfert 2 galaxies that showed double peaks in their SDSS\footnote{Sloan Digital Sky Survey \citep{York00}.} narrow emission line spectra and were detected in the Very Large Array (VLA) FIRST\footnote{Faint Images of the Radio Sky at Twenty-Centimeters \citep{Becker95}.} survey; the latter being a criteria for the authors to propose for Very Long Baseline Array (VLBA) observations. Double peak emission line AGN are potential candidates for binary black hole systems \citep{Zhang07}.

\begin{figure}
\centering{\includegraphics[width=7.3cm]{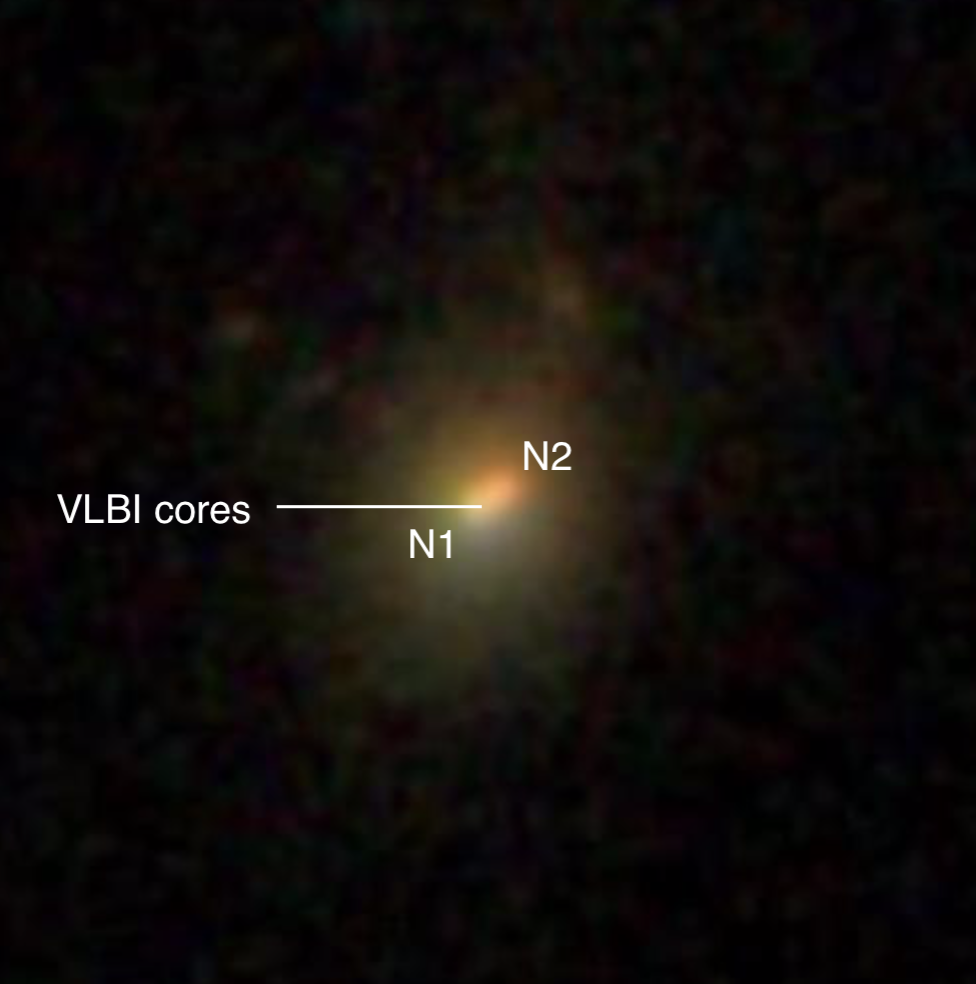}}
\caption{\small The color composite SDSS image of KISSR\,102 showing the position of the detected VLBI cores, coincident with the south-eastern optical nucleus N1.}
\label{fig1}
\end{figure}

KISSR\,102 is hosted by an elliptical galaxy at a redshift of 0.066320 (Figure~\ref{fig1}). At this distance, 1~milliarcsec (mas) corresponds to a linear extent of 1.237 parsec\footnote{Cosmology: H$_0$ = 73~km~s$^{-1}$~Mpc$^{-1}$, $\Omega_{m}$ = 0.27, $\Omega_{v}$ = 0.73}. The SDSS and Pan-STARRS\footnote{Panoramic Survey Telescope and Rapid Response System \citep{ChambersMMF16}.} optical images of KISSR\,102 show the presence of two nuclei (Figure~\ref{fig4}); the brighter reddish nucleus (referred to hereafter as N1) is located closer to the galactic photo-centre and is coincident with the radio AGN, while N2 is a bluish nucleus located 1\farcs2 (=1.54 kpc at the distance of KISSR\,102) to the north-west of N1. 

This source was identified as an ``offset AGN'' candidate by \citet{Comerford14} because its AGN emission lines are offset in the line-of-sight velocity from the stellar absorption lines by $\approx$200~km~s$^{-1}$. A radio point source with a flux density of 11.3 mJy at 1.4 GHz is detected in  KISSR\,102 in its VLA FIRST (beam~$5.4\arcsec=6.7$~kpc) image. The SDSS spectrum shows the tentative presence of double peaks in the H$\beta$ line, but not in other lines. Throughout this paper, the spectral index $\alpha$ is defined such that the flux density at frequency $\nu$ is $S_\nu\propto\nu^\alpha$.

\section{Radio Data Analysis}
We observed KISSR\,102 with all ten antennas of the VLBA in a phase-referencing experiment at 4.9 and 1.5~GHz, on December 26, 2018, and January 06, 2019, respectively {(Project IDs: BK219A, BK219B)}. However, data from the Brewster antenna was unusable at 1.5~GHz, resulting effectively in 9 antennas for the experiment at that frequency. All data were acquired at an aggregate bit rate of 2048 Mbits~s$^{-1}$ with dual polarization, 16 channels per subband, each of bandwidth 32~MHz and an integration time of 2 seconds. The compact calibrator 1246+285 (1.78$^\circ$ away from KISSR\,102) with a small (x,~y) positional uncertainty of (0.31,~0.47) mas, was used as the phase reference calibrator. The target and the phase reference calibrator were observed in a ``nodding'' mode in a 5~min cycle (2~min on calibrator and 3~min on source) for good phase calibration. After including scans of the fringe-finder 3C\,345 and the phase-check source 1502+291, the experiment lasted a total of 8 hours at each frequency.

\begin{figure*}
\centering{
\includegraphics[width=6.75cm]{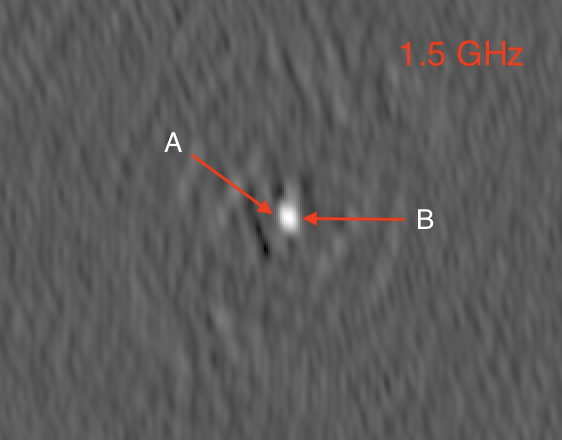}
\includegraphics[width=6.85cm]{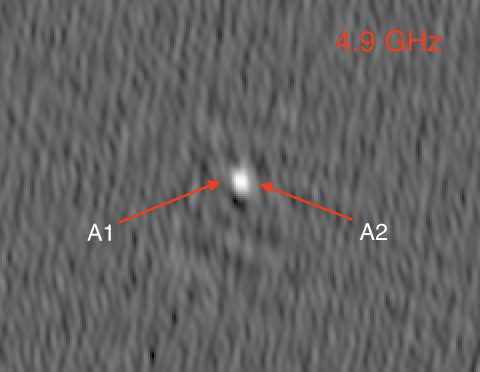}
\includegraphics[width=7.2cm]{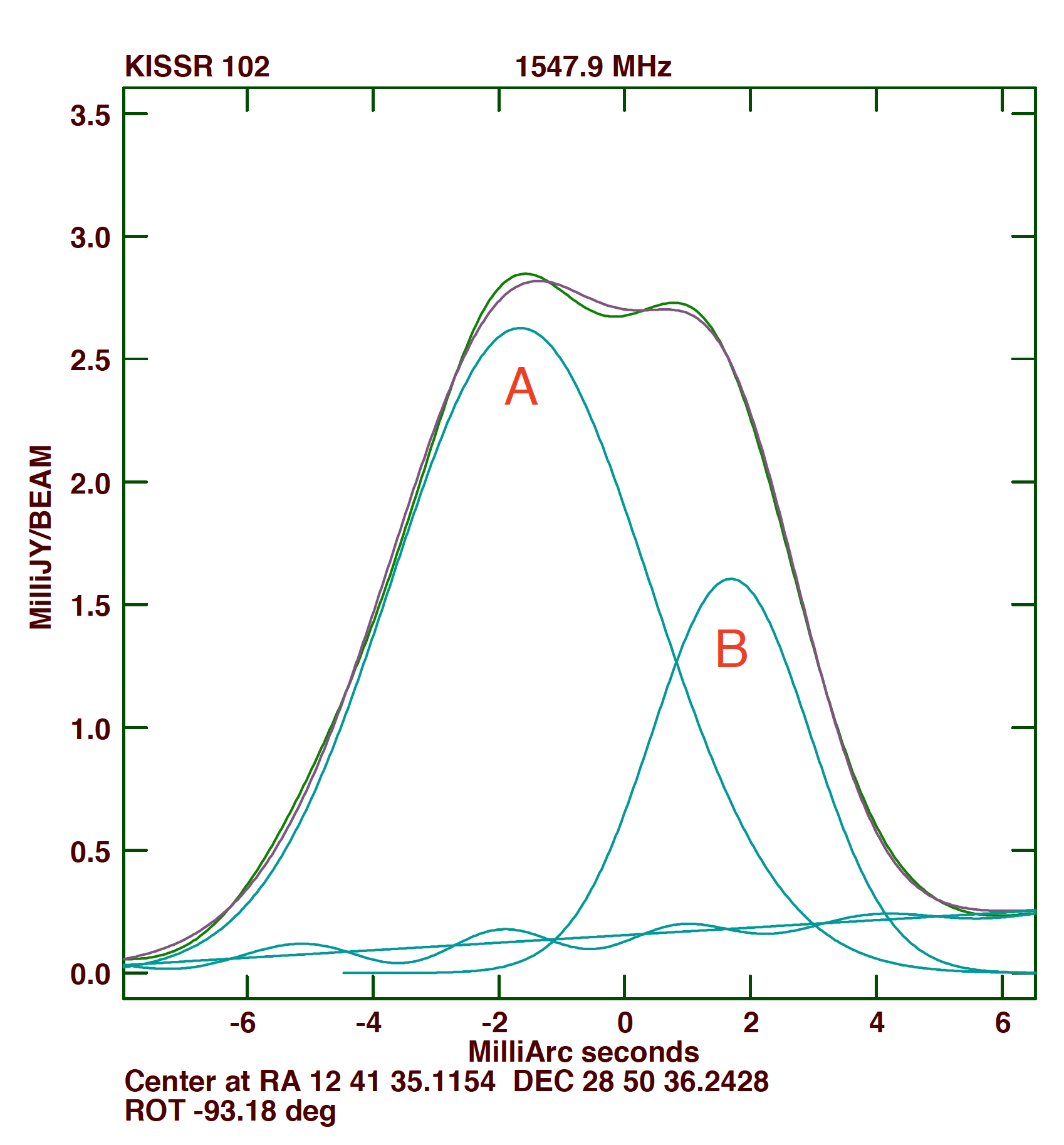}
\includegraphics[width=7.2cm]{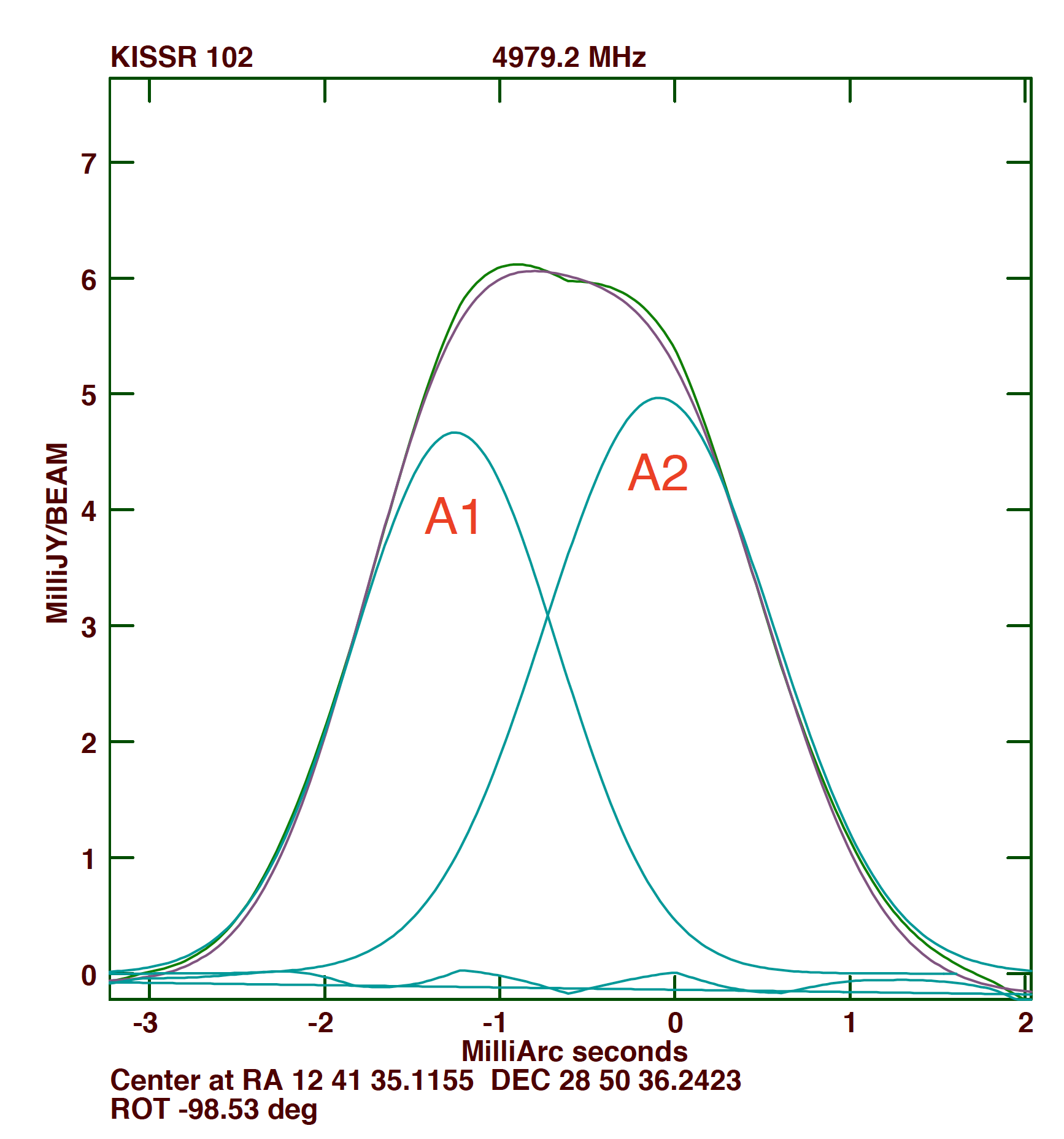}}
\caption{\small Top left \& right: 1.5 and 4.9 GHz VLBA images of KISSR\,102 in grey-scale showing double radio cores at a projected separation at 4.8 parsec (A, B) and 1.7 parsec (A1, A2), respectively. {Bottom left \& right: Two Gaussian component fits to intensity cuts (at PA = $-93.2\degr$ at 1.5~GHz and $-98.5\degr$ at 4.9~GHz) where the green, purple lines are from the observed data, fitted model, respectively. }The two Gaussian components along with the fitted continuum base (straight line) and residuals (wiggly line) are shown in blue.}
\label{fig2}
\end{figure*}

\begin{figure*}
\centering{
\includegraphics[width=8cm,trim=90 520 170 50]{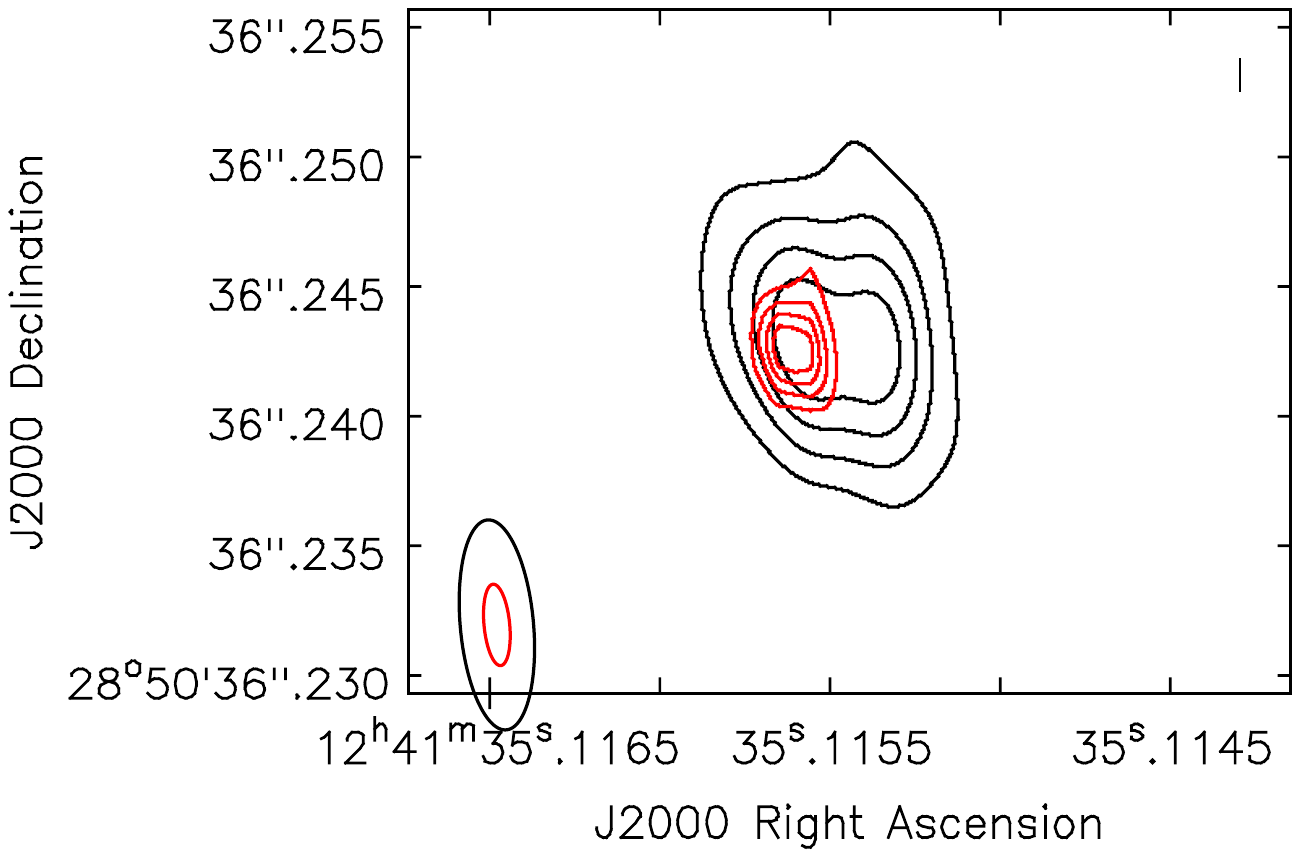}
\includegraphics[width=8cm,trim=40 480 220 100]{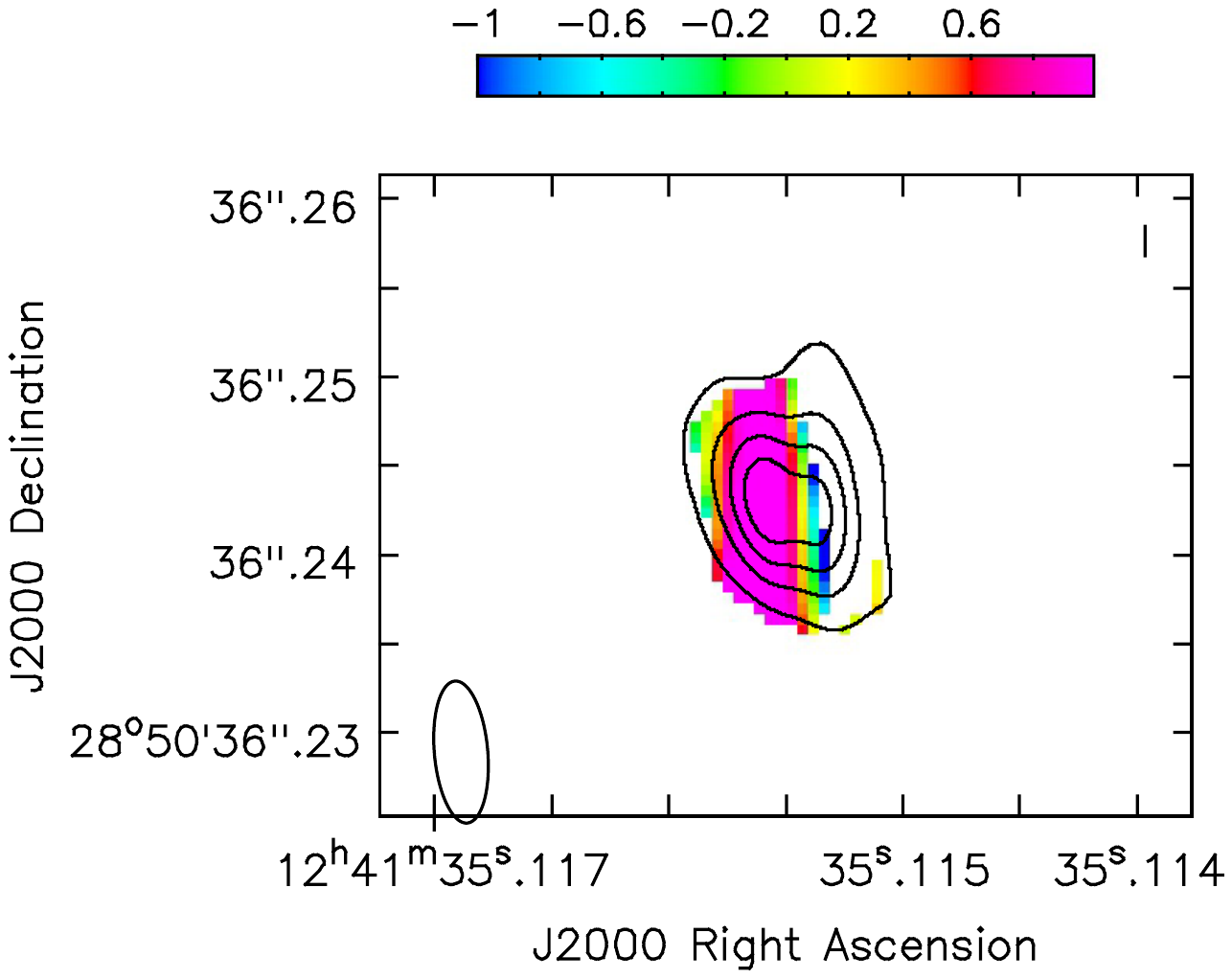}}
\caption{\small {Left \& right: Black 1.5~GHz and red 4.9~GHz contours superimposed on each other. Contour levels are 0.2, 0.4, 0.6, 0.8 of 2.9~mJy~beam$^{-1}$ at 1.5 GHz, 0.2, 0.4, 0.6, 0.8 of 6.0~mJy~beam$^{-1}$ at 4.9~GHz. The beam sizes are $8.1\times2.8$~mas at PA = 5.2$\degr$ at 1.5~GHz, $3.1\times0.9$~mas at PA = 5.5$\degr$ at 4.9~GHz. The $1.5-4.9$~GHz spectral index image in colour, made with images of beam $8.0\times3.0$~mas at PA = 5.0$\degr$, superimposed by 1.5~GHz radio contours in black with levels at 0.2, 0.4, 0.6, 0.8 of 3.0~mJy~beam$^{-1}$.}}
\label{fig3}
\end{figure*}

The data were reduced in AIPS using standard calibration procedures as described in the AIPS Cookbook\footnote{\url{http://www.aips.nrao.edu/CookHTML/CookBookap3.html\#x179-387000C}}. The source was detected with an offset of (360, 27)~mas from the centre of the image; {the task UVFIX was used to bring the target to the centre}. The final uniformly-weighted images (made with a ROBUST parameter of $-5$) are presented in Figures~\ref{fig2} and \ref{fig3}. The final {\it rms} noise in the images is {$\sim90~\mu$Jy~beam$^{-1}$ at 1.5~GHz and $\sim75~\mu$Jy~beam$^{-1}$ at 4.9~GHz. The restoring beams at 1.5 and 4.9~GHz are ($8.1\times2.8$~mas$^2$ at PA\footnote{All position angles are measured in degrees East of North.}$=5.2\degr$) and ($3.1\times0.9$~mas$^2$ at PA$=5.5\degr$), respectively.} The flux density values as well as the positions quoted in this paper have been obtained by using the Gaussian-fitting AIPS task JMFIT. The $1.5-4.9$~GHz spectral index image was made after convolving both frequency images with a beam of {$8.0\times3.0$~mas, PA $=5\degr$ and the AIPS task COMB}; pixels with intensity values below three times the {\it rms} noise were blanked at both frequencies. The spectral index image is presented in the bottom panel of Figure~\ref{fig2}.

\section{Optical data analysis}
{Upon visual inspection of the optical image present in the SDSS and Pan-STARRS1 databases, the host galaxy does not show any evident signs of recent mergers or strong interactions with nearby galaxies, appearing rather as a featureless elliptical.} Two objects, perhaps galaxies, are located approximately 10$\arcsec$ and 14$\arcsec$ from the galaxy centre, but the physical distance between the members of the triplet is unknown. The galaxy shows, however, a peculiar color gradient and what seems like a double nucleus. The brighter nucleus N1 is at a projected separation of 1.54 kpc from the nucleus N2. 

We used \textsc{galfit} \citep[v3.0.5,][]{GALFIT2002}, to model the Pan-STARRS1 DR2 
\textit{i}-band image. To derive the point spread function (PSF), we fitted the following model to the bright star located at (r,$\theta$) = ($77^{\prime\prime}$, $-138^{\circ}$) from N1:
\begin{align*}
f(z) &= A \times (1 + kz + z^{1.67})^{-1},\\ 
z & = \left(\frac{x^2}{2\sigma^{2}_{xx}} + \frac{y^2}{2\sigma^{2}_{yy}} + \sigma_{xy}xy\right),
\end{align*}
\noindent
obtaining $A = 7.8 \times 10^{5}$, $k=0.487$, $\sigma_{xx} = 2.38$ pixels, $\sigma_{yy} = 2.3$ pixels, $\sigma_{xy} = -0.01$ pixels. A model PSF was built with $A=1$, and provided to \textsc{galfit} as an input parameter.

\begin{figure*}
\begin{center}$
\begin{array}{cc}
\includegraphics[trim=15 30 0 0, scale=0.4]{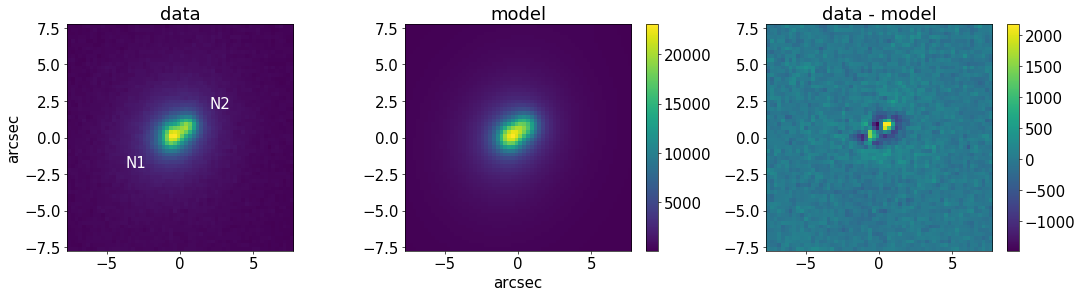} \\
\end{array}$
\end{center}
\caption{\small From left to right: PanSTARRS1 \textit{i}-band image of KISSR\,102 (zoomed on the nucleus), \textsc{galfit} model (two point sources on top of a Sersic profile and a sky background), and the corresponding residual.}
\label{fig4}
\end{figure*}

An area of 24$\arcsec$ in radius, centered on the galaxy, was fitted with a sky background, a Sersic profile for the host galaxy, a Sersic profile for the faint nearby galaxy at (r,$\theta$) $ = (11^{\prime\prime}, -24^{\circ}$), and a point source for the faint object at (r,$\theta$) $ = (14^{\prime\prime}, 56^{\circ}$). To fit the nucleus we attempted a number of approaches, namely: two point sources, two point sources with an edge-on disk, one point source and an edge-on disk. Furthermore, point sources were fitted using as a template the semi-empirical model specified above, a Moffat profile, or a nearby isolated star. The best result is obtained with the semi-empirical PSF, and the adoption of two nuclear point sources. Models including an edge-on disk did not converge to a solution. A zoom-in on the nucleus, best fitting model, and residuals are shown in Figure~\ref{fig4}. The residual image in the rightmost panel of Figure~\ref{fig4} casts doubt on the presence of two point sources in the galaxy, showing that N2 is poorly modelled. At visual inspection it is clear that the morphology of the nucleus varies with wavelength: N2 is brighter at shorter wavelengths ($g$-band) where the nucleus appears boxy and the light centroid falls in between N1 and N2. At longer wavelengths, however, N1 becomes more prominent and determines the light centroid. We used the \textsc{IRAF} task \textsc{ellipse} \citep{Jed87} to perform an isophotal decomposition of the $i$-band PanSTARRS image, finding that N1 is coincident to within 0\farcs18 (223 parsec, or 0.7 pixels) with the intensity-weighted center defined by the inner isophotes, that is with semi-major axis in the range $2\farcs5 \leq r \leq 10^{\prime\prime}$ ($3 \leq r \leq 12$ kpc).

\begin{figure*}
\centering{
\includegraphics[width=14cm,trim=0 0 0 180]{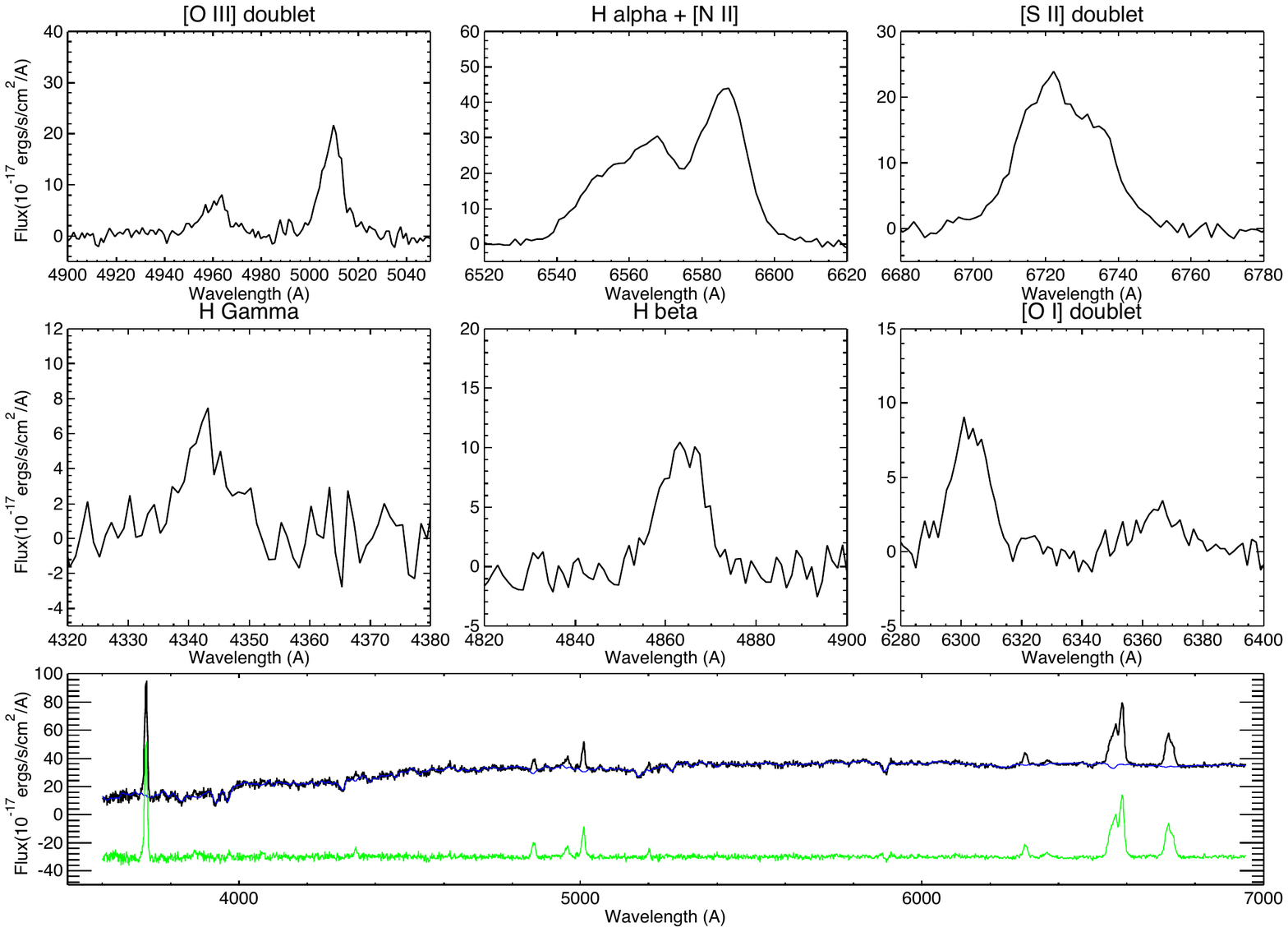}
\includegraphics[width=17.5cm,trim=0 0 0 360]{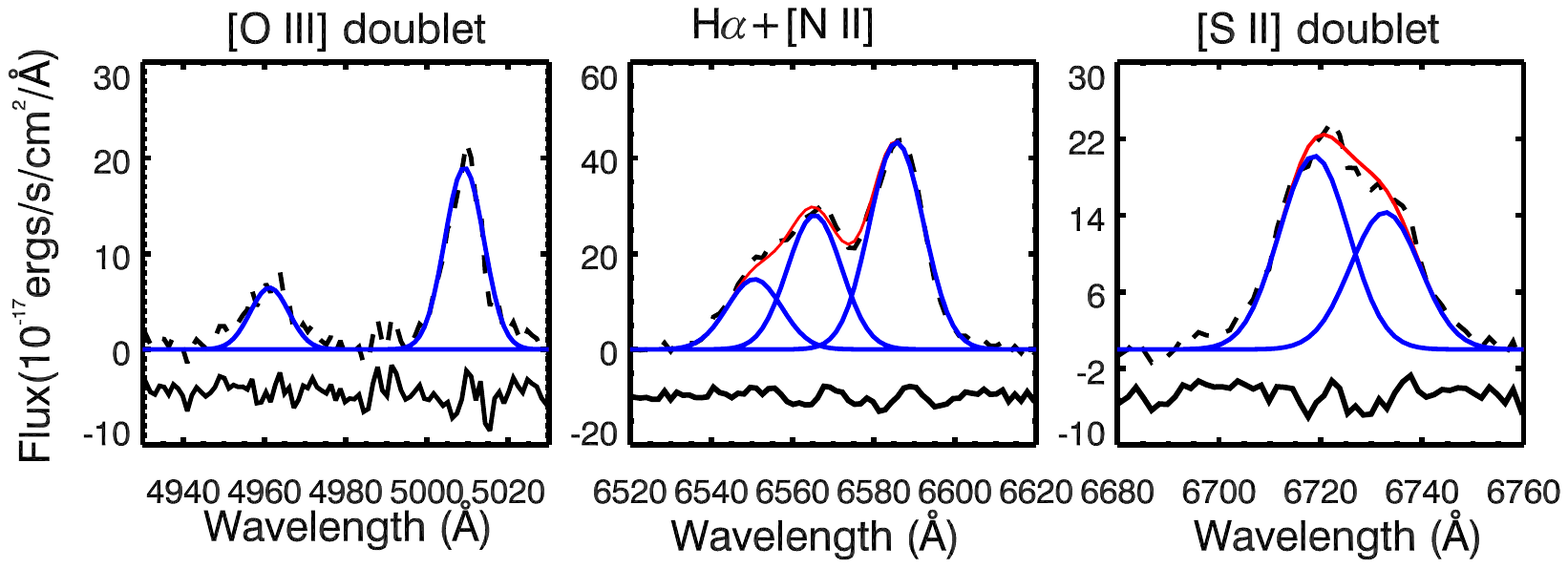}
\includegraphics[width=17.5cm,trim=0 0 0 670]{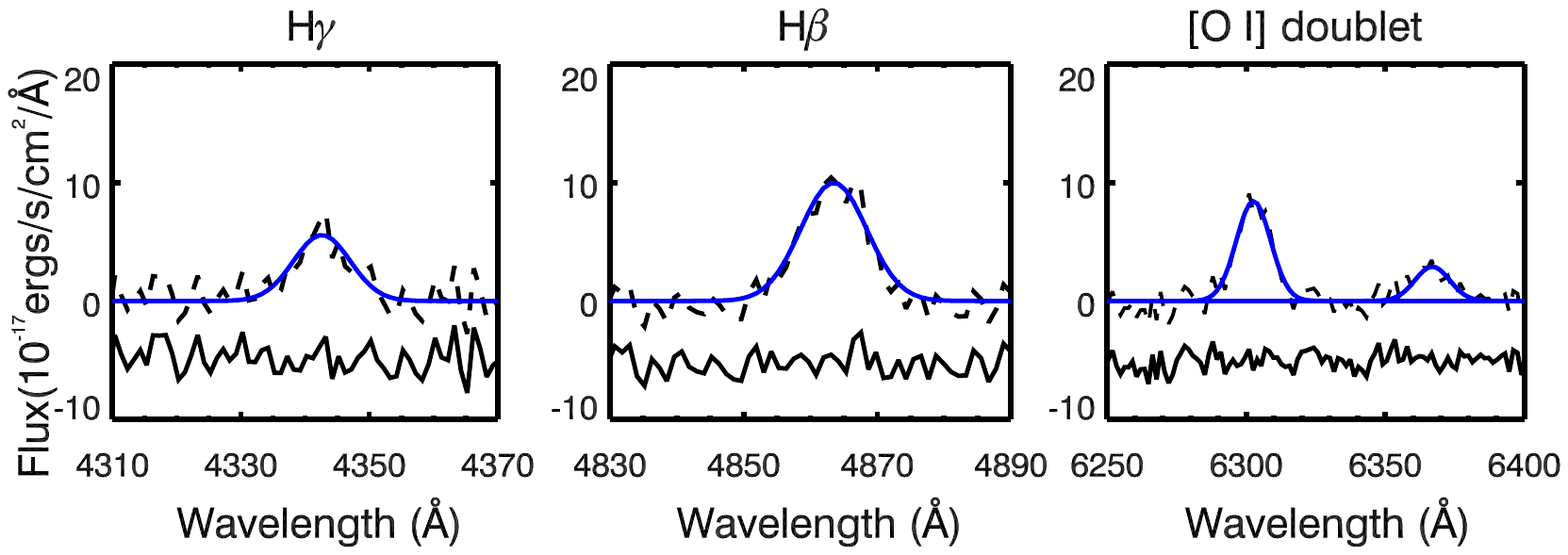}}
\vspace{-13.5cm}
\caption{\small SDSS optical spectrum fitted with Gaussian components. Only the H$\beta$ line shows a weak double peak. A higher resolution spectrum is needed to observe double peaks in other emission lines.}
\label{fig5}
\end{figure*}

We fitted the SDSS spectrum\footnote{SDSS spectra are acquired through a fiber of diameter $3\arcsec$, or 3.7~kpc at the distance of KISSR\,102} using the pPXF (Penalized Pixel-Fitting stellar kinematics extraction) code of \citet{Cappellari04,Cappellari17}. Details of the line-fitting procedure have been presented earlier in \citet{Kharb17b,Kharb19}. The spectrum was corrected for reddening using the E(B$-$V) value from \citet{Schlegel98}. The observed stellar velocity dispersion was $340\pm10$~km~s$^{-1}$. A single Gaussian component was adequate to describe the [S II] lines which are used as a template for other narrow lines. No broad component was detected in the H$\alpha$ line, making this a Type 2 LINER galaxy. There were no signatures of an outflow component in [O III]. The results of the analysis are shown in Figure~\ref{fig5}. The reddening-corrected spectra are shown in black and the best fit model is over-plotted in blue. The best fit model was subtracted from the de-reddened observed spectrum to obtain the pure emission line spectrum, shown in the middle panel in green. IDL programs which use the {\tt MPFIT} function for non-linear least-square optimisation were used to fit the emission line profiles with Gaussian functions and to obtain the best fit parameters and the associated errors which are reported in Table~\ref{tabprop}.

\begin{figure*}
\centering{
\includegraphics[width=8.6cm,trim=0 0 0 150]{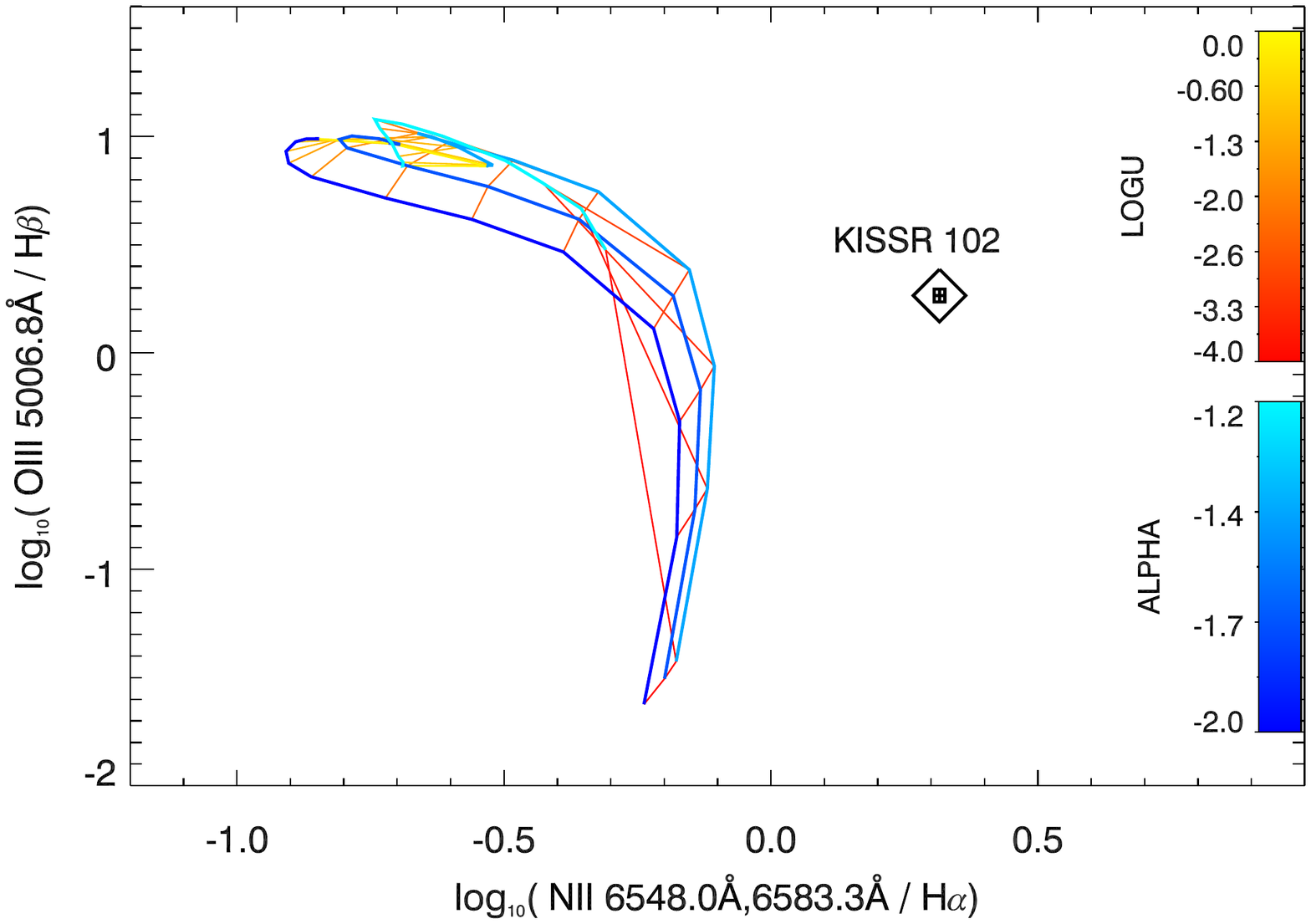}
\includegraphics[width=8.6cm,trim=0 0 0 150]{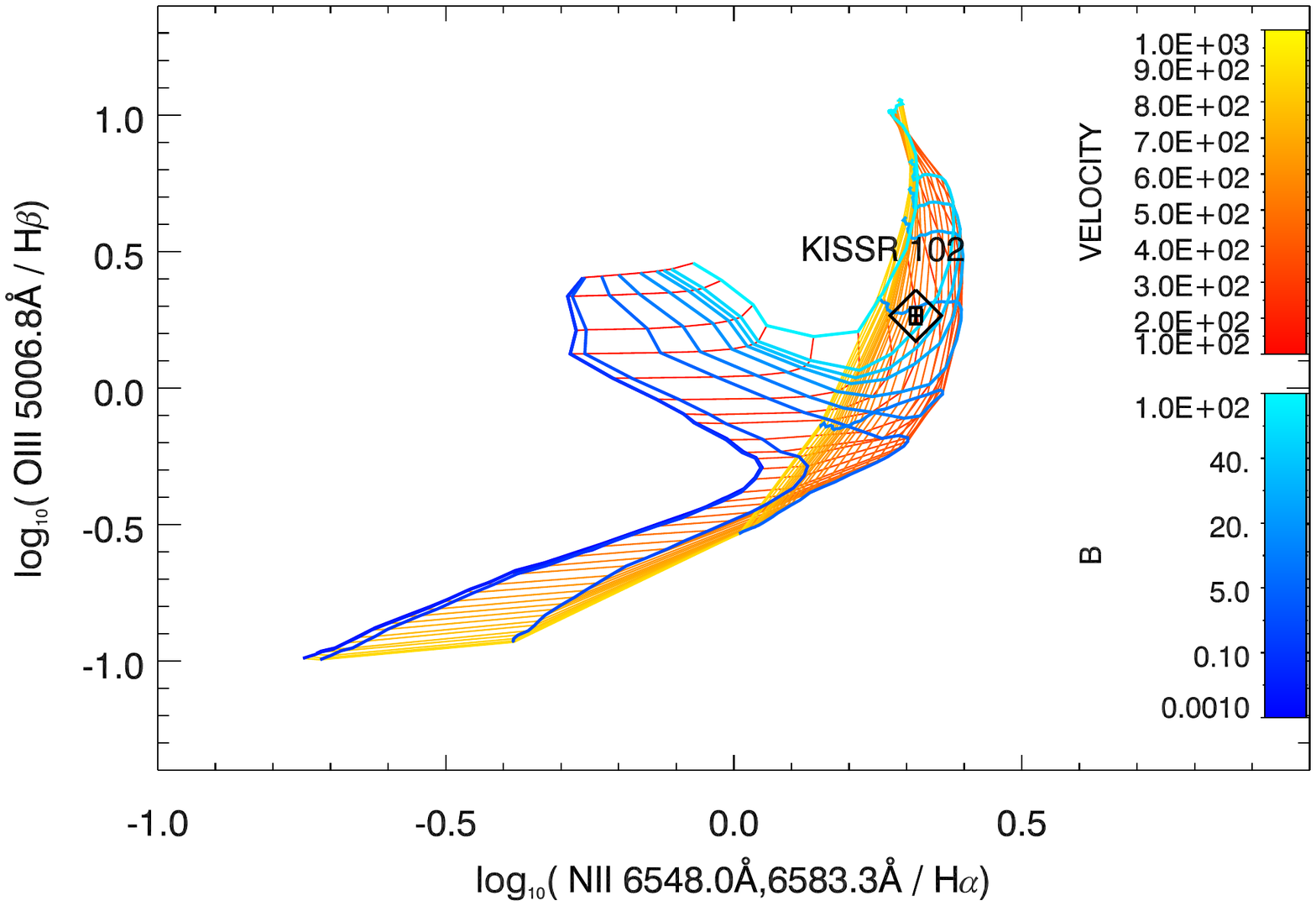}
\includegraphics[width=8.6cm,trim=0 0 0 400]{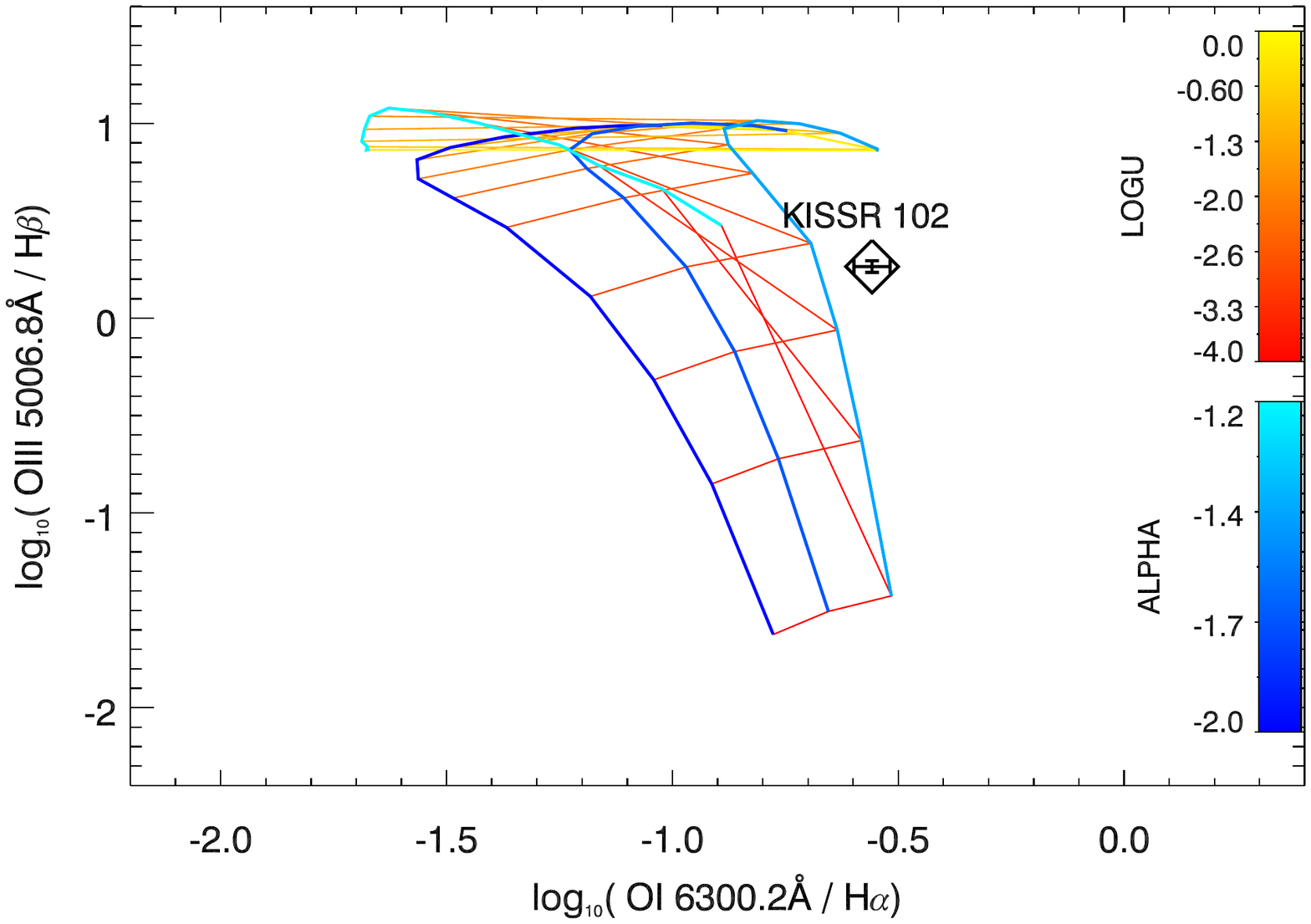}
\includegraphics[width=8.6cm,trim=0 0 0 400]{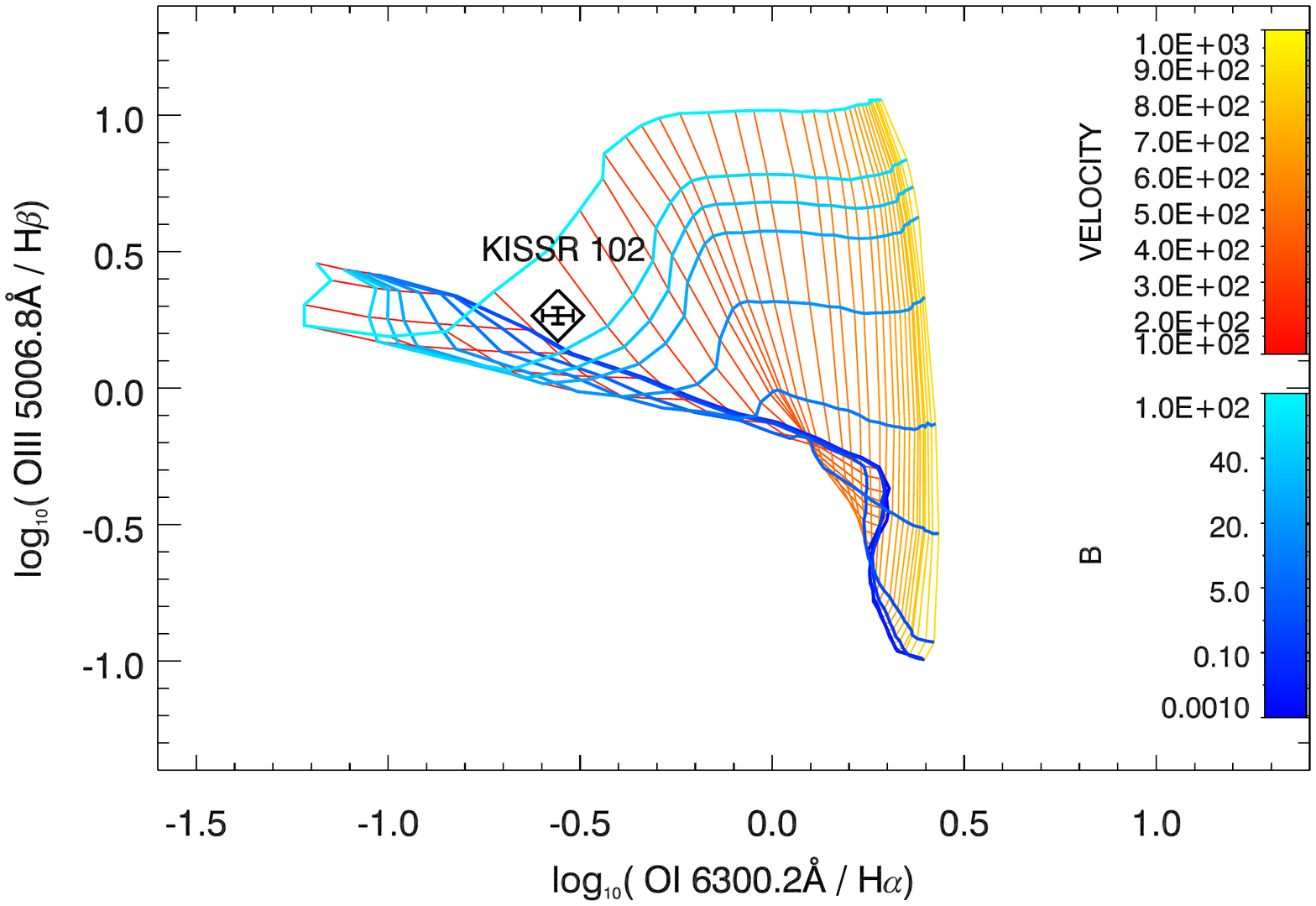}
\includegraphics[width=8.6cm,trim=0 0 0 400]{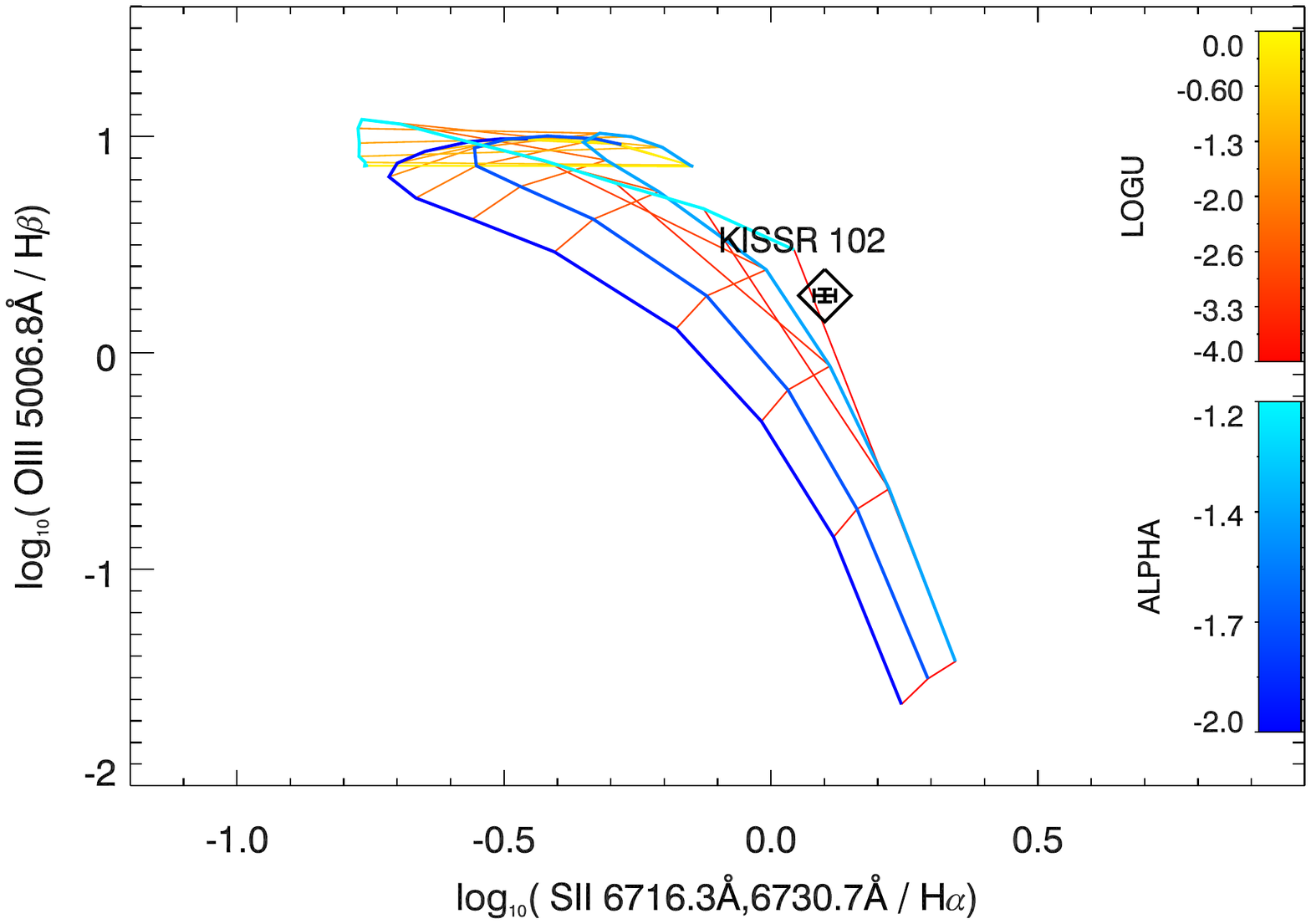}
\includegraphics[width=8.6cm,trim=0 0 0 400]{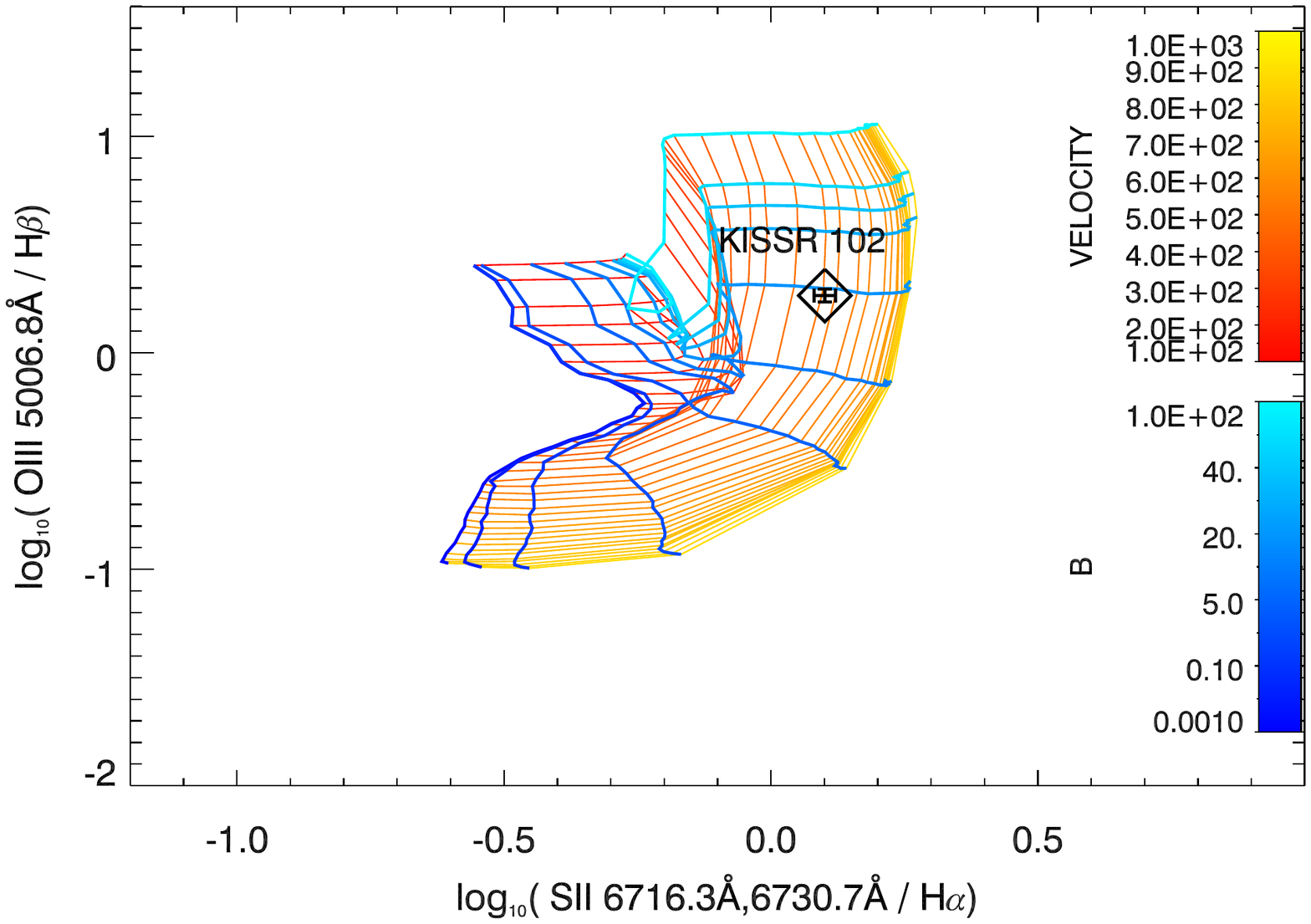}}
\vspace{-2.5cm}
\caption{Results from the MAPPINGS III code suggest that shock-heating (the three right panels) rather than AGN photoionisation (the three left panels) is responsible for the observed line ratios in KISSR\,102.}
\label{fig6}
\end{figure*}

Finally, the MAPPINGS III shock and photoionization modelling code \citep{Dopita1996,Allen2008} was used to predict the line ratios in order to match the data in KISSR\,102. The IDL Tool for Emission-line Ratio Analysis \citep[ITERA;][]{Groves2010} was used for generating the line ratio diagrams. In Figure~\ref{fig6} we show the optical line ratio diagrams for AGN photoionisation and shock-only models. These models are run for a density of 100~cm$^{-3}$ and solar metallicity. The shock-only model, with velocities in the range  $200-300$~km~s$^{-1}$, produces line ratios consistent with the observed [N II]/H$\alpha$ and [O I]/H$\alpha$ line ratios. However, a higher shock velocity of $\sim$600 km~s$^{-1}$ is needed to reproduce the [S II]/H$\alpha$ line ratio. The AGN photoionization model under-predicts the observed line ratios. In order to reconcile the AGN model with the observed data, either a higher density or a higher metallicity is required. The possibility of higher densities can be ruled out however, by the high [S II]/H$\alpha$ line ratio observed in this source. This therefore implies that the shock-only model is more credible for the case of KISSR\,102.

\section{Results}
The 1.5~GHz VLBA image reveals two radio components (A and B in Figure~\ref{fig2}) at a projected separation at 4.8~parsec. The 4.9~GHz image detects only component A, but seems to again show two barely-resolved radio components at a projected separation at 1.7~parsec (components A1, A2 in Figure~\ref{fig2}). We note that the radio AGN is coincident with the optical nucleus N1 (see Figure~\ref{fig1}). {The peak surface brightness of components A and B are $\sim$2.7~mJy~beam$^{-1}$ and $\sim$1.7~mJy~beam$^{-1}$, respectively. Their total flux densities are $\sim$4.4~mJy~beam$^{-1}$ and $\sim$2.6~mJy~beam$^{-1}$, respectively. The task JMFIT was not able to provide unique constraints on the flux densities of components A1 and A2 at 4.9~GHz. We therefore relied on the intensity-slice Gaussian-fitting task SLFIT to obtain the peak intensities of the two Gaussian components, which were $\sim$4.7~mJy~beam$^{-1}$ for A1 and $\sim$5.0~mJy~beam$^{-1}$ for A2. We fixed these for running JMFIT and obtained component sizes, which were subsequently used to obtain the brightness temperatures.} The bottom panels of Figure~\ref{fig2} show the two Gaussian components fitted to intensity cuts across the radio source at both frequencies.

The $1.5-4.9$~GHz spectral index image in Figure~\ref{fig2} shows that the mean spectral index value of component A is inverted at {$+0.64\pm0.08$.} This is consistent with component A being the self-absorbed base of a radio jet, like in radio-powerful AGN. Using five times the {\it rms} noise in the 4.9~GHz image, the $1.5-4.9$~GHz spectral index limit for component B is very steep with $\alpha\lesssim-1.6$, consistent with ``aged'' optically thin synchrotron emission. Alternatively, component B could be radio emission from a second black hole belonging to a low luminosity AGN; steep-spectrum radio cores are often found in low luminosity AGN \citep[][and references therein]{Deane14,Fu15,Kharb15b}. 

{We note that the ``implied'' steep spectrum of component B could in principle also mean that it is diffuse emission that gets resolved out at 5~GHz. This is reminiscent of the single radio component that was detected by \citet{Kharb15b} in KISSR\,1494, which was not centrally concentrated in emission and got resolved out in images with different weighting schemes. \citet{Kharb15b} suggested that component in KISSR\,1494 resembled the base of a possible coronal wind, rather than an AGN jet component on mas-scales, which is typically more compact. If this is also true for component B, then it makes it further consistent with the binary black hole picture.}

Using the relation from \citet{Ulvestad05} for unresolved radio components, the brightness temperatures of components A and B {at 1.5~GHz are $\sim2.0\times10^8$~K and $\sim2.5\times10^8$~K, respectively. Similarly, the brightness temperatures of components A1 and A2 at 4.9 GHz are $\sim1.1\times10^9$~K and $\sim4.0\times10^8$~K, respectively.} These high brightness temperatures are consistent with emission originating in AGN components \citep{Falcke00}.

The black hole mass based on the $M-\sigma$ relation \citep{McConnell13} is $1.7\times10^9$~M$_\sun$. The star-formation rate (SFR) from the H$\alpha$ narrow line luminosity ($\sim3.8 \times10^{40}$~erg~s$^{-1}$) and the \citet{Kennicutt98} relation is a modest  $0.30\pm0.01~M_\sun$~yr$^{-1}$. Clearly, any ongoing merger activity in KISSR\,102, has not resulted in a boosted SFR. The bolometric luminosity based on the [O III] line luminosity and the scaling relation from \citet{Heckman04} is $\sim6.4\times10^{43}$ erg~s$^{-1}$. The Eddington luminosity is $2.1\times10^{47}$~erg~s$^{-1}$, while the Eddington ratio is extremely low at $3\times10^{-4}$. This low estimate is however, consistent with what is observed in LINER nuclei \citep{Ho08}.

\section{Discussion}
We present two different scenarios to explain the multi-band data of KISSR\,102, and discuss their pros and cons along with possible implications.

\subsection{Scenario 1: A \& B are from a single AGN}
The inverted spectral index and high brightness temperature of radio component A at 1.5~GHz is consistent with it being the synchrotron self-absorbed base of an AGN jet. Component B, at a projected distance of $4.8$~parsec from component A and with a steep spectral index, could be an optically thin jet component belonging to this AGN. Components A1 and A2, which are coincident with component A, and are at a projected separation of $1.7$~parsec from each other, could also be a ``core-jet'' structure. In this scenario, the 5~GHz data are further resolving the 1.5~GHz radio component A. 

\subsubsection{Scenario 1: Pros}
This scenario can explain the line ratios as being from a fast jet that is shock-heating the NLR clouds, in tandem with the putative recoiling AGN. That is, the required shock velocities of $\sim$600~km~s$^{-1}$ needed to explain the observed line ratios, could be a combination of the AGN recoil velocity ($\sim200$~km~s$^{-1}$) and a fast radio jet along the same direction. 

The double-peaked H$\beta$ emission line, which although has a weak signature due to the noisy SDSS spectrum, could be explained as a result of jet-NLR interaction \citep[e.g.,][]{Kharb17b,Kharb19}. Detecting double peaks in other {(or any)} emission lines would require a higher resolution optical spectrum, similar to what has been observed in the ``offset AGN'' CXO J101527.2+625911 by \citet{Kim19} using Keck spectra. 
\subsubsection{Scenario 1: Cons}
The spectral indices of components A and B are drastically different. In the case of a ``core-jet'' structure from a single AGN, a more gradual change in the spectral indices would be expected \citep[e.g.,][]{Kharb09}. If instead, A and B are the mini-lobes of a compact symmetric object (CSO), their spectral index values should be similar \citep[e.g.,][]{Argo15}.

Moreover, a ``core-jet'' structure requires the jet to be highly relativistic so that the counter-jet emission is Doppler-dimmed. Using the peak surface brightness of component B and the noise in the 1.5~GHz image along the counter-jet direction, the jet-to-counterjet surface brightness ratio is $R_J\sim20$. Assuming the jet structural parameter to be $p=3.6$ \citep[i.e., a continuous jet with $\alpha=-1.6$;][]{Urry95}, and the jet orientation w.r.t. line of sight to be greater than or equal to $\sim50\degr$ \citep[i.e., torus half opening angle;][]{Simpson96} for this type 2 AGN, we derive a lower limit on the jet speed to be $\sim0.6c$. It is unclear if radio-quiet LINERs can harbour such highly relativistic radio jets. However, LINER nuclei do show an inverse correlation between Eddington accretion rate and (X-ray) radio-loudness \citep{Ho08}. Therefore, a relativistic jet in KISSR\,102 cannot be ruled out in the absence of multi-epoch VLBI data that can determine jet speeds. 

{If the relatively brighter component A2 is the radio core and the relatively dimmer component A1 is jet emission, then this jet is actually in the counter-jet direction based on components A and B. If true, this would be more consistent with Scenario 2, where components A and B are two separate AGN. However, the caveat is that components A1 and A2 are not fully resolved, and their flux densities are not uniquely constrained (Section~4). This makes it difficult to confirm
the core-jet direction at 5~GHz with present data.}

\subsubsection{Scenario 1: Implications}
Strong interaction between the radio jet and surrounding inhomogenous media that can result in a drastic spectral steepening of the jet \citep[e.g.,][]{vanBreugel84}, is needed for the ``core-jet'' picture to be feasible. Scenario 1 could therefore imply strong jet-medium interaction; the dense surrounding media could also be responsible for the synchrotron self-absorption of the counterjet emission, alleviating the need for large Doppler jet speeds on parsec-scales.

\subsection{Scenario 2: A \& B are a binary AGN}
In this scenario, both components A and B are individual AGN at a projected separation of 4.8~parsec. The inverted radio spectrum of component A as well the steep spectrum of component B are both consistent with originating in LINER or Seyfert nuclei \citep{Ho08}. Here, components A1 and A2 would be a ``core-jet'' structure associated with the AGN at position A. In addition, there would be a third black hole coincident with the optical nucleus N2.

\subsubsection{Scenario 2: Pros}
The projected separation of $\sim5$~parsec between components A and B is roughly the ``hard'' binary separation for a black hole binary with a total mass of $\sim10^9$~M$_\sun$. A supermassive black hole binary is deemed to be ``hard'' \citep{Merritt13} if its separation is less than $\sim a_h$ (in parsec), where 
\begin{eqnarray*}
a_h = \frac{G\mu}{4\sigma^2} \approx 0.27 \left(1+q\right)^{-1} \left(\frac{M_2}{10^7 M_\odot}\right)
\left(\frac{\sigma}{200\ \mathrm{km}\ \mathrm{s}^{-1}}\right)^{-2}.
\end{eqnarray*}

Here, $q\equiv M_2/M_1$ and $\mu\equiv M_1M_2/(M_1+M_2)$, where $M_1$ and $M_2$ are the two black hole masses. Assuming $M_1+M_2=1.7\times10^9 M_\odot$, $\sigma=340$ km s$^{-1}$, and $q=1$ yields $a_h\approx7.9$~parsec; {broadly} consistent with the projected separation of A and B. 

At the separation $\sim a_h$, a massive binary inside a galactic stellar nucleus is expected to ``stall''. This is because the binary has already ejected all stars intersecting its orbit and capable of carrying away orbital energy \citep{Merritt13}. In the case of KISSR\,102, the orbital hardening could have been aided by a three-body interaction due to the presence of three supermassive black holes. This would be consistent with the presence of the second optical nucleus N2 and KISSR\,102's classification as an ``offset AGN''. We discuss this in more detail in Section~\ref{sec:imp}. The double-peaked H$\beta$ line, albeit weak, could be an indicator of a binary black hole with individual, unmerged NLRs \citep[whose extents can span a $\sim$kpc;][]{Schmitt03b} in KISSR\,102. Higher resolution optical spectra are needed to detect double peaks in other ({or any}) emission lines. Finally, the three-body interaction could have created shocks in the emission-line gas surrounding the binary black hole system, which can explain the MAPPINGS line ratio results in KISSR\,102.

\subsubsection{Scenario 2: Cons}
A triple supermassive black hole system is required as the initial condition. Confirmed triple supermassive black hole systems are rare \citep{Deane14,Pfeifle19}. Moreover, the three black hole configuration would probably be short lived, so that we seem to be catching it at a very special time.

\subsubsection{Scenario 2: Implications}\label{sec:imp}
Since an ``offset AGN" could be the result of an asymmetric release of gravitational radiation due to a binary black hole coalescence, or due to a gravitational slingshot recoil kick involving three black holes, we speculate that there were initially three supermassive black holes in the center of the host galaxy of KISSR\,102. The three-body interaction ejected the lightest of the three black holes, which now resides inside the optical nucleus N2. This in turn ``hardened'' the orbit of the remaining two black holes, creating a close binary system, at the position of the nucleus N1. The combined mass of the three black holes could explain the large {inferred mass of} $\sim10^9$~M$_\sun$ in this galaxy; {emission line} signatures of these cannot be observed in the SDSS spectra because the SDSS fiber covers the entire region containing the nuclei N1 and N2.

We estimate the geodetic precession period of the potential binary black hole system using the relation, $P_{prec}\sim600~{r_{16}}^{5/2}(M/m)~{M_8}^{-3/2}$~yr \citep{Begelman80}. Here, $r_{16}$ is the binary separation in units of $10^{16}$~cm, (M/m) is the mass ratio of primary to secondary black hole (assumed to be unity) and $M_{8}$ is the mass of the primary supermassive black hole in units of $10^{8}~M_{\odot}$. Using $M_8\sim17$ and $r_{16}\sim1478$, we estimate the geodetic precession period to be $\sim7.2\times10^8$~yr.

\subsection{Could N1 or N2 be an Infalling Dwarf Galaxy?}
Could the optical nuclei N1 be a infalling dwarf galaxy moving towards the galactic center, similar to what is observed in the ``offset AGN", NGC\,3341 \citep{Bianchi13}. Similar to KISSR\,102, NGC\,3341 has a line-of-sight velocity that is (blue-)shifted by 200~km~s$^{-1}$ relative to the host galaxy nucleus, which lies at a projected separation of 9.5$\arcsec$ = 5.2~kpc from the AGN. \citet{Bianchi13} find that the Seyfert 2 AGN hosted by a dwarf galaxy and detected at radio frequencies, is infalling towards the LINER/HII region at the centre of the larger primary galaxy, and this infall is triggering the Seyfert activity. Because of the large mass ratio of the merging galaxies, star-formation is mostly triggered in the dwarf galaxy due to tidal forces but not much in the primary galaxy \citep[e.g., as observed in the simulations of][]{Cox08}. A scenario similar to that taking place in NGC\,3341 could explain the low SFR observed in the primary galaxy of KISSR\,102. 

However, no {obvious} signatures of an ongoing merger, like the presence of tidal tails, shells or asymmetric isophotes, are observed in KISSR\,102. WISE colours are used to identify heavily obscured AGN; these are identified with colors $W1-W2\ge=0.8$ \citep{Stern05}. For KISSR\,102, $W1-W2=0.1$ indicating that the nuclear regions are not dust obscured. No IR emission is detected in the WISE W3 and W4 images. All these point to one thing: {there is} no hot dust heated by AGN continuum and no warm/cold dust heated by young stellar population. Moreover, as the dwarf galaxy would be nearly coincident with the photometric center of the galaxy (within 0.18$\arcsec$ or 223 parsec), it seems unrealistic that the dwarf galaxy fails to get disrupted before reaching the host galaxy center. As the optical nucleus N2 could be fitted with a Sersic profile, could it be an infalling dwarf galaxy  instead? The current optical data are not particularly constraining to confirm or reject this hypothesis.

\section{Summary and Conclusions}
We report the detection of an intriguing parsec-scale radio source in the kinematically offset AGN, KISSR\,102. The host galaxy of KISSR\,102 includes two optical nuclei at a projected separation of 1.54~kpc, N1 and N2, to the south-east and north-west, respectively. Phase-referenced VLBA observations at 1.5 and 4.9 GHz of this LINER galaxy, have detected double radio components (components A \& B) at the projected separation of 4.8~parsec at 1.5~GHz, and a partially-resolved double radio structure at 4.9~GHz that is coincident with component A at 1.5~GHz; these radio detections are at the position of the optical nucleus, N1. The brightness temperatures of all the detected radio components are high, $\gtrsim10^8$~K, consistent with them being components of a radio AGN. The $1.5-4.9$~GHz spectral index is inverted {($\alpha\sim+0.64\pm0.08$)} for component A and steep for component B ($\alpha \lesssim-1.6$). The sharp change between the spectral index values of A and B is inconsistent with it being a ``core-jet'' structure from a single AGN, or the mini-lobes of a CSO; a more gradual spectral index change would be expected in the former case, while a similarity in spectral indices would be required in the latter. However, strong jet-medium interaction with dense surrounding media could be invoked to reconcile with the core-jet scenario. Alternatively, there could be a parsec-scale-separated binary AGN in KISSR\,102, which could be the end result of a three-body interaction involving three supermassive black holes.

\begin{table*}
\scriptsize
\caption{Fitted Line Parameters for KISSR\,102}
\begin{center}
\begin{tabular}{lclcllc}
\hline\hline
{Line} & {$\lambda_{0}$} & {$\lambda_{c}\pm$error} & {$\Delta\lambda\pm$error} & {$f_{p}\pm$error} & {$F\pm$error}& {$L\pm$error}\\
{(1)}   & {(2)}  & {(3)}                 & {(4)} & {(5)} & {(6)} & {(7)}\\ \hline
$[\mathrm {S~II}]$ & 6718.3  & 6718.71$\pm$0.35  &   6.74$\pm$0.09 &   20.22$\pm$0.74  & 341.78$\pm$13.28  &    2.82$\pm$0.11 \\
$[\mathrm {S~II}]$ & 6732.7  & 6732.72$\pm$0.35  &   6.76$\pm$0.01 &   14.32$\pm$0.73  & 242.71$\pm$12.33  &   2.00$\pm$0.10 \\
$[\mathrm {N~II}]$ & 6549.9  & 6550.58$\pm$0.12  &   6.57$\pm$0.01 &   14.67$\pm$0.23  & 241.71$\pm$3.79   &  2.00$\pm$0.03 \\
$[\mathrm {N~II}]$ & 6585.3  & 6585.60$\pm$0.12  &   6.61$\pm$0.00 &   43.29$\pm$0.67  & 716.97$\pm$11.15  &   5.92$\pm$0.09 \\
H$\alpha$         & 6564.6  & 6565.42$\pm$0.20  &   6.59$\pm$0.01 &  28.05$\pm$0.64   & 463.09$\pm$10.64  &   3.82$\pm$0.09 \\
H$\beta$          & 4862.7  & 4863.48$\pm$0.33  &   4.82$\pm$0.01 &   10.00$\pm$0.48  & 120.88$\pm$5.87   &  1.00$\pm$0.05 \\
$[\mathrm {O~III}]$& 4960.3  & 4961.23$\pm$0.18  &   4.63$\pm$0.17 &    6.44$\pm$0.22  &  74.68$\pm$3.78   &  0.62$\pm$0.03 \\
$[\mathrm {O~III}]$& 5008.2  & 5009.19$\pm$0.18  &   4.68$\pm$0.04 &   18.99$\pm$0.65  & 222.55$\pm$7.86   &  1.84$\pm$0.06 \\
H$\gamma$         & 4341.7  & 4342.62$\pm$0.56  &   4.28$\pm$0.01 &    5.57$\pm$0.50  &  59.73$\pm$5.41   &  0.49$\pm$0.04 \\
$[\mathrm {O~I}]$  & 6302.0  & 6302.62$\pm$0.37  &   6.08$\pm$0.36 &    8.42$\pm$0.48  & 128.29$\pm$10.53  &   1.06$\pm$0.09 \\
$[\mathrm {O~I}]$  & 6365.5  & 6366.93$\pm$0.38  &   6.14$\pm$0.06 &    2.90$\pm$0.37  &  44.58$\pm$5.68   &  0.37$\pm$0.05 \\
\hline
\end{tabular}
\end{center}
{Column~1: Emission lines that were fitted with Gaussian components. Column~2: Rest wavelength in vacuum in $\AA$. Columns~3, 4: Central wavelength and line width ($\sigma$) in $\AA$ along with respective errors. Column~5: Peak line flux in units of $10^{-17}$~ergs~cm$^{-2}$~s$^{-1}~\AA^{-1}$ with error. (All line flux densities have been corrected for galactic or foreground reddening.) Column~6: Total line flux in $10^{-17}$~ergs~cm$^{-2}$~s$^{-1}~\AA^{-1}$. Column~7: Line luminosity in units of $10^{40}$~ergs~s$^{-1}$. }
\label{tabprop}
\end{table*}

\acknowledgments
The National Radio Astronomy Observatory is a facility of the National Science Foundation operated under cooperative agreement by Associated Universities, Inc. SS acknowledges support from the Science and Engineering Research Board, India through the Ramanujan Fellowship. 

\vspace{5mm}
\facilities{VLBA, Sloan}
{\software{pPXF \citep{Cappellari04,Cappellari17}, MAPPINGS III \citep{Dopita1996,Allen2008}, GALFIT \citep[v3.0.5,][]{Peng02}, IRAF \citep{Tody93}, AIPS \citep{vanMoorsel96}}}

\bibliographystyle{aasjournal}
\bibliography{ms}

\end{document}